\documentstyle[amssymb,floats,aps,prb,epsf]{revtex}
\epsfclipon
\begin{document}
\twocolumn[\hsize\textwidth\columnwidth\hsize\csname
@twocolumnfalse\endcsname

\draft
\title{The charge asymmetry in superconductivity of hole- and
electron-doped cuprates}
\author{Tianxing Ma and Shiping Feng}
\address{Department of Physics, Beijing Normal University, Beijing
100875, China}
\maketitle
\begin{abstract}
Within the $t$-$t'$-$J$ model, the charge asymmetry in
superconductivity of hole- and electron-doped cuprates is studied
based on the kinetic energy driven superconducting mechanism. It
is shown that superconductivity appears over a narrow range of
doping in electron-doped cuprates, and the superconducting
transition temperature displays the same kind of the doping
controlled behavior that is observed in the hole-doped case.
However, the maximum achievable superconducting transition
temperature in the optimal doping in electron-doped cuprates is
much lower than that of the hole-doped case due to the
electron-hole asymmetry.
\end{abstract}

]
\bigskip

\narrowtext

The parent compounds of cuprate superconductors are believed to
belong to a class of materials known as Mott insulators with the
antiferromagnetic (AF) long-range order (AFLRO), then
superconductivity occurs by the electron or hole doping
\cite{kastner,bednorz,tallon,tokura}. Both hole-doped and
electron-doped cuprate superconductors have the layered structure
of the square lattice of the CuO$_{2}$ plane separated by
insulating layers \cite{kastner,tokura}. It has been found from
experiments that only an approximate symmetry in the phase diagram
exists about the zero doping line between electron- and hole-doped
cuprates \cite{hasan,sawa}. For hole-doped cuprates
\cite{kastner,bednorz,tallon}, AFLRO disappears rapidly with
doping, and is replaced by a disordered spin liquid phase, then
the systems become superconducting (SC) over a wide range of the
hole doping concentration $\delta$, around the optimal $\delta\sim
0.15$, however, AFLRO survives until superconductivity appears
over a narrow range of $\delta$ around the optimal $\delta\sim
0.15$ in electron-doped cuprates, where the maximum achievable SC
transition temperature is much lower than hole-doped cuprates
\cite{tokura,tokura1,peng}. Although this electron-hole asymmetry
is observed in the phase diagram \cite{kastner,hasan,sawa}, the
charge carrier Cooper pairs in the both optimally electron- and
hole-doped cuprate superconductors have a dominated d-wave
symmetry \cite{tsuei,martindale,armitage,tsuei1}. Since the strong
electron correlation is common for both hole-doped and
electron-doped cuprates, many of the physical properties of
electron-doped cuprates resemble that of the hole-doped case. By
virtue of systematic studies using the nuclear magnetic resonance,
and muon spin rotation techniques, particularly the inelastic
neutron scattering, it has been well established that the AF
short-range correlation in both hole- and electron-doped cuprate
superconductors coexists with the SC state
\cite{kastner,yamada,yamada1,kang}. These provide a clear link
between the charge carrier pairing mechanism and magnetic
excitations, and show that both hole- and electron-doped cuprate
superconductors have similar underlying SC mechanism
\cite{yamada2}.

Within the $t$-$J$ model, we \cite{feng1,feng2} have discussed the
kinetic energy driven SC mechanism in hole-doped cuprates based on
the charge-spin separation (CSS) fermion-spin theory \cite{feng3},
where the dressed holons interact occurring directly through the
kinetic energy by exchanging dressed spin excitations, leading to
a net attractive force between dressed holons, then the electron
Cooper pairs originating from the dressed holon pairing state are
due to the charge-spin recombination, and their condensation
reveals the SC ground-state \cite{feng1}. The SC transition
temperature is controlled by both charge carrier gap function and
single particle coherent weight. This single particle coherent
weight grows linearly with increasing doping in the underdoped and
optimally doped regimes, and then decreases with increasing doping
in the overdoped regime, which leads to that the maximal
superconducting transition temperature occurs around the optimal
doping, and then decreases in both underdoped and overdoped
regimes \cite{feng2}. In this paper, we study the charge asymmetry
in superconductivity of hole- and electron-doped cuprates along
with this line. We show that superconductivity appears over a
narrow range of doping in electron-doped cuprates, and the maximum
achievable SC transition temperature in the optimal doping is
lower than that of the hole-doped case due to the electron-hole
asymmetry.

In both hole- and electron-doped cuprates, the characteristic
feature is the presence of the two-dimensional CuO$_{2}$ plane
\cite{kastner,tokura} as mentioned above, and it seems evident
that the unusual behaviors are dominated by this plane. Although
the $t$-$J$ model captures the essential physics of the doped
CuO$_{2}$ plane \cite{anderson}, the electron-hole asymmetry may
be accounted for by including further neighbor hoping $t'$
\cite{hybertson}. Therefore we start from the $t$-$t'$-$J$ model,
\begin{eqnarray}
H&=&-t\sum_{i\hat{\eta}\sigma}C^{\dagger}_{i\sigma}
C_{i+\hat{\eta}\sigma}+t'\sum_{i\hat{\tau}\sigma}
C^{\dagger}_{i\sigma}C_{i+\hat{\tau}\sigma}\nonumber \\
&+&\mu\sum_{i\sigma}
C^{\dagger}_{i\sigma}C_{i\sigma}+J\sum_{i\hat{\eta}}{\bf S}_{i}
\cdot {\bf S}_{i+\hat{\eta}},
\end{eqnarray}
with $\hat{\eta}=\pm\hat{x},\pm\hat{y}$, $\hat{\tau}=\pm\hat{x}
\pm\hat{y}$, $C^{\dagger}_{i\sigma}$ ($C_{i\sigma}$) is the
electron creation (annihilation) operator, ${\bf S}_{i}=
C^{\dagger}_{i}{\vec\sigma}C_{i}/2$ is spin operator with
${\vec\sigma}=(\sigma_{x},\sigma_{y},\sigma_{z})$ as Pauli
matrices, and $\mu$ is the chemical potential. For the electron
doping, we can perform a particle-hole transformation $C_{i\sigma}
\rightarrow C^{\dagger}_{i-\sigma}$, so that the difference
between hole and electron doping is expressed as the sign
difference of the hopping parameters, i.e., $t>0$ and $t'>0$ for
hole doping and $t<0$ and $t'<0$ for electron doping \cite{feng5},
then the $t$-$t'$-$J$ model (1) in both hole- and electron-doped
cases is always subject to an important on-site local constraint
to avoid the double occupancy, i.e., $\sum_{\sigma}
C^{\dagger}_{i\sigma}C_{i\sigma} \leq 1$. This single occupancy
local constraint can be treated properly within the CSS
fermion-spin theory \cite{feng3}, where the constrained electron
operators are decoupled as, $C_{i\uparrow}=
h^{\dagger}_{i\uparrow} S^{-}_{i}$, $C_{i\downarrow}=
h^{\dagger}_{i\downarrow}S^{+}_{i}$, with the spinful fermion
operator $h_{i\sigma}= e^{-i\Phi_{i\sigma}}h_{i}$ describes the
charge degree of freedom together with some effects of the spin
configuration rearrangements due to the presence of the doped
charge carrier itself (dressed charge carrier), while the spin
operator $S_{i}$ describes the spin degree of freedom (dressed
spin), then the electron local constraint for the single occupancy
is satisfied in analytical calculations, and low-energy behavior
of the $t$-$t'$-$J$ model (1) in this CSS fermion-spin
representation can be expressed as \cite{feng1,feng2,feng3},
\begin{eqnarray}
H&=&-t\sum_{i\hat{\eta}}(h_{i\uparrow}S^{+}_{i}
h^{\dagger}_{i+\hat{\eta}\uparrow}S^{-}_{i+\hat{\eta}}+
h_{i\downarrow}S^{-}_{i}h^{\dagger}_{i+\hat{\eta}\downarrow}
S^{+}_{i+\hat{\eta}})\nonumber\\
&+&t'\sum_{i\hat{\tau}}(h_{i\uparrow}S^{+}_{i}
h^{\dagger}_{i+\hat{\tau}\uparrow}S^{-}_{i+\hat{\tau}}+
h_{i\downarrow}S^{-}_{i}h^{\dagger}_{i+\hat{\tau}\downarrow}
S^{+}_{i+\hat{\tau}}) \nonumber \\
&-&\mu\sum_{i\sigma}h^{\dagger}_{i\sigma}h_{i\sigma}+J_{{\rm eff}}
\sum_{i\hat{\eta}}{\bf S}_{i}\cdot {\bf S}_{i+\hat{\eta}},
\end{eqnarray}
with $J_{{\rm eff}}=(1-x)^{2}J$, and $\delta=\langle
h^{\dagger}_{i\sigma}h_{i\sigma}\rangle=\langle h^{\dagger}_{i}
h_{i}\rangle$ is the doping concentration. As a consequence, the
kinetic energy terms in the $t$-$t'$-$J$ model have been expressed
as the interactions between dressed charge carriers and spins,
which reflects that even the kinetic energy terms in the
$t$-$t'$-$J$ Hamiltonian have strong Coulombic contributions due
to the restriction of no doubly occupancy of a given site. These
interactions between dressed charge carriers and spins are quite
strong, and we \cite{feng1,feng2} have shown that in the case
without AFLRO, these interactions can induce the dressed charge
carrier pairing state (then the electron Cooper pairing state) by
exchanging dressed spin excitations in the higher power of the
doping concentration $\delta$. Since the SC state in both hole-
and electron-doped cuprates is characterized by the electron
Cooper pairs, forming SC quasiparticles \cite{tsuei,tsuei1}, and
in the real space the gap function and pairing force have a range
of one lattice spacing \cite{shen5}, therefore the order parameter
for the electron Cooper pair can be expressed as, $\Delta=\langle
C^{\dagger}_{i\uparrow}C^{\dagger}_{i+\hat{\eta}\downarrow}-
C^{\dagger}_{i\downarrow}C^{\dagger}_{i+\hat{\eta}\uparrow}
\rangle=\langle h_{i\uparrow}h_{i+\hat{\eta}\downarrow}S^{+}_{i}
S^{-}_{i+\hat{\eta}}-h_{i\downarrow}h_{i+\hat{\eta}\uparrow}
S^{-}_{i} S^{+}_{i+\hat{\eta}}\rangle =-\langle S^{+}_{i}
S^{-}_{i+\hat{\eta}}\rangle\Delta_{h}$, with the dressed charge
carrier pairing order parameter $\Delta_{h}=\langle
h_{i+\hat{\eta}\downarrow}h_{i\uparrow}-h_{i+\hat{\eta}\uparrow}
h_{i\downarrow}\rangle$, which shows that the SC order parameter
is closely related to the dressed charge carrier pairing
amplitude, and is proportional to the number of charge carriers,
and not to the number of electrons. Following the Eliashberg's
strong coupling theory \cite{eliashberg}, we obtain the
self-consistent equations that satisfied by the full dressed
charge carrier diagonal and off-diagonal Green's functions as
\cite{feng1},
\begin{mathletters}
\begin{eqnarray}
g(k)&=&g^{(0)}(k)+g^{(0)}(k)[\Sigma^{(h)}_{1}(k)g(k)\nonumber \\
&-&\Sigma^{(h)}_{2}(-k)\Im^{\dagger}(k)], \\
\Im^{\dagger}(k)&=&g^{(0)}(-k)[\Sigma^{(h)}_{1}(-k)
\Im^{\dagger}(-k) \nonumber \\
+\Sigma^{(h)}_{2}(-k)g(k)],
\end{eqnarray}
\end{mathletters}
respectively, where the four-vector notation $k=({\bf k},
i\omega_{n})$, the dressed charge carrier mean-field (MF) diagonal
Green's function \cite{feng1} $g^{(0)-1}(k)= i\omega_{n}-
\xi_{{\bf k}}$, the MF dressed charge carrier excitation spectrum
$\xi_{{\bf k}}= Zt\chi_{1}\gamma_{{\bf k}}-Zt'\chi_{2}
\gamma_{{\bf k}}'-\mu$, with $\gamma_{{\bf k}}=(1/Z)
\sum_{\hat{\eta}}e^{i{\bf k}\cdot\hat{\eta}}$, $\gamma_{{\bf k}}'
=(1/Z)\sum_{\hat{\tau}}e^{i{\bf k} \cdot\hat{\tau}}$, $Z$ is the
number of the nearest neighbor or second-nearest neighbour sites,
the dressed spin correlation functions $\chi_{1}=\langle S_{i}^{+}
S_{i+\hat{\eta}}^{-}\rangle$ and $\chi_{2}=\langle S_{i}^{+}
S_{i+\hat{\tau}}^{-}\rangle$, and the dressed charge carrier
self-energy functions \cite{feng1,feng2},
\begin{mathletters}
\begin{eqnarray}
\Sigma^{(h)}_{1}(k)&=&{1\over N^{2}}\sum_{{\bf p,p'}}(Zt
\gamma_{{\bf p+p'+k}}-Zt'\gamma_{{\bf p+p'+k}}')^{2}\nonumber\\
&\times&{1\over\beta}\sum_{ip_{m}}g(p+k){1\over\beta}
\sum_{ip'_{m}}D^{(0)}(p')D^{(0)}(p'+p), \\
\Sigma^{(h)}_{2}(k)&=&{1\over N^{2}}\sum_{{\bf p,p'}}(Zt
\gamma_{{\bf p+p'+k}}-Zt'\gamma_{{\bf p+p'+k}}')^{2} \nonumber\\
&\times& {1\over\beta}\sum_{ip_{m}}\Im(-p-k){1\over\beta}
\sum_{ip'_{m}} D^{(0)}(p') D^{(0)}(p'+p),
\end{eqnarray}
\end{mathletters}
where $p=({\bf p},ip_{m})$, $p'=({\bf p'},ip_{m}')$, and the MF
dressed spin Green's function \cite{feng1},
\begin{eqnarray}
D^{(0)}(p)={B_{{\bf p}}\over (ip_{m})^{2}-\omega_{{\bf p}}^{2}},
\end{eqnarray}
with $B_{{\bf p}}= \lambda_{1}[2\chi^{z}_{1}(\epsilon\gamma_{{\bf
p}}-1)+ \chi_{1}(\gamma_{{\bf p}}-\epsilon)]-\lambda_{2}
(2\chi^{z}_{2} \gamma_{{\bf p}}'-\chi_{2})$, $\lambda_{1}=
2ZJ_{eff}$, $\lambda_{2} =4Z\phi_{2}t'$, $\epsilon=1+2t\phi_{1}
/J_{{\rm eff}}$, and the MF dressed spin excitation spectrum
$\omega^{2}_{{\bf p}}=A_{1} (\gamma_{{\bf k}})^{2}+ A_{2}
(\gamma'_{{\bf k}})^{2}+ A_{3} \gamma_{{\bf k}}\gamma'_{{\bf k}}
+A_{4}\gamma_{{\bf k}}+ A_{5} \gamma'_{{\bf k}}+A_{6}$, with
$A_{1}=\alpha\epsilon \lambda_{1}^{2}(\epsilon \chi^{z}_{1}+
\chi_{1}/2)$, $A_{2}= \alpha \lambda_{2}^{2}\chi^{z}_{2}$,
$A_{3}=-\alpha\lambda_{1} \lambda_{2}(\epsilon \chi^{z}_{1}+
\epsilon\chi^{z}_{2}+ \chi_{1}/2)$, $A_{4}=-\epsilon
\lambda_{1}^{2}[\alpha (\chi^{z}_{1}+\epsilon\chi_{1}/2)+(\alpha
C^{z}_{1}+ (1-\alpha)/(4Z)-\alpha\epsilon\chi_{1}/(2Z))+(\alpha
C_{1}+ (1-\alpha)/(2Z)-\alpha\chi^{z}_{1}/2)/2]+\alpha\lambda_{1}
\lambda_{2}(C_{3}+\epsilon\chi_{2})/2$, $A_{5}=-3\alpha
\lambda^{2}_{2}\chi_{2}/(2Z)+\alpha\lambda_{1}\lambda_{2}
(\chi^{z}_{1}+\epsilon\chi_{1}/2+C^{z}_{3})$, $A_{6}=
\lambda^{2}_{1}[\alpha C^{z}_{1}+(1-\alpha)/(4Z)-\alpha\epsilon
\chi_{1}/(2Z)+\epsilon^{2}(\alpha C_{1}+(1-\alpha)/(2Z)-\alpha
\chi^{z}_{1}/2)/2]+\lambda^{2}_{2}(\alpha C_{2}+(1-\alpha)/(2Z)-
\alpha\chi^{z}_{2}/2)/2)-\alpha\epsilon\lambda_{1}\lambda_{2}
C_{3}$, and the dressed charge carrier's particle-hole parameters
$\phi_{1}=\langle h^{\dagger}_{i\sigma} h_{i+\hat{\eta}\sigma}
\rangle$, $\phi_{2}=\langle h^{\dagger}_{i\sigma}
h_{i+\hat{\tau}\sigma}\rangle$, the dressed spin correlation
functions $\chi^{z}_{1}=\langle S_{i}^{z}S_{i+\hat{\eta}}^{z}
\rangle$, $\chi^{z}_{2}=\langle S_{i}^{z}S_{i+\hat{\tau}}^{z}
\rangle$, $C_{1}=(1/Z^{2})\sum_{\hat{\eta},\hat{\eta'}}\langle
S_{i+\hat{\eta}}^{+}S_{i+\hat{\eta'}}^{-}\rangle$, $C^{z}_{1}=
(1/Z^{2})\sum_{\hat{\eta},\hat{\eta'}}\langle S_{i+\hat{\eta}}^{z}
S_{i+\hat{\eta'}}^{z}\rangle$, $C_{2}=(1/Z^{2})
\sum_{\hat{\tau},\hat{\tau'}}\langle S_{i+\hat{\tau}}^{+}
S_{i+\hat{\tau'}}^{-}\rangle$, $C_{3}=(1/Z)\sum_{\hat{\tau}}
\langle S_{i+\hat{\eta}}^{+} S_{i+\hat{\tau}}^{-}\rangle$, and
$C^{z}_{3}=(1/Z) \sum_{\hat{\tau}}\langle S_{i+\hat{\eta}}^{z}
S_{i+\hat{\tau}}^{z}\rangle$. In order to satisfy the sum rule of
the correlation function $\langle S^{+}_{i}S^{-}_{i}\rangle=1/2$
in the case without AFLRO, the important decoupling parameter
$\alpha$ has been introduced in the MF calculation \cite{feng1},
which can be regarded as the vertex correction.

In Eq. (4), the self-energy function $\Sigma^{(h)}_{2}(k)$ is
called as the effective dressed charge carrier gap function since
it contains both pairing force and dressed charge carrier gap
function, while the self-energy function $\Sigma^{(h)}_{1}(k)$
renormalizes the MF dressed charge carrier spectrum, and therefore
it describes the single particle (quasiparticle) coherence. In
particular, $\Sigma^{(h)}_{2}(k)$ is an even function of
$i\omega_{n}$, while $\Sigma^{(h)}_{1}(k)$ is not. For the
convenience of discussions, we separate $\Sigma^{(h)}_{1}(k)$ into
its symmetric and antisymmetric parts as, $\Sigma^{(h)}_{1}(k)=
\Sigma^{(h)}_{1e}(k) +i\omega_{n} \Sigma^{(h)}_{1o}(k)$, then
$\Sigma^{(h)}_{1e}(k)$ and $\Sigma^{(h)}_{1o}(k)$ are both even
functions of $i\omega_{n}$. According to the Eliashberg's strong
coupling theory \cite{eliashberg}, we can define the charge
carrier single particle (quasiparticle) coherent weight
$Z^{-1}_{F}(k)=1- \Sigma^{(h)}_{1o}(k)$. On the other hand, the
retarded function ${\rm Re} \Sigma^{(h)}_{1e}(k)$ may be a
constant, independent of (${\bf k}, \omega$). It just renormalizes
the chemical potential, and therefore can be neglected
\cite{eliashberg}. Furthermore, we only study the static limit of
the effective dressed charge carrier gap function and single
particle coherent weight, i.e.,
$\Sigma^{(h)}_{2}(k)=\bar{\Delta}_{h}({\bf k})$, and
$Z^{-1}_{F}({\bf k})=1-\Sigma^{(h)}_{1o}({\bf k})$. Although
$Z_{F}({\bf k})$ still is a function of ${\bf k}$, the wave vector
dependence is unimportant, since everything happens at the
electron Fermi surface. As in the previous discussions within the
$t$-$J$ model \cite{feng2}, the special wave vector can be
estimated qualitatively from the electron momentum distribution as
${\bf k}_{0}={\bf k_{A}}-{\bf k_{F}}$ with ${\bf k_{A}}= [\pi,
\pi]$ and ${\bf k_{F}}\approx [(1-x)\pi/2, (1-x)\pi/2]$, which
guarantees $Z_{F}=Z_{F} ({\bf k}_{0})$ near the electron Fermi
surface. In this case, the dressed charge carrier diagonal and
off-diagonal Green's functions in Eq. (3) can be rewritten
explicitly as,
\begin{mathletters}
\begin{eqnarray}
g(k)&=&{1\over 2}\left (1+ {\bar{\xi_{{\bf k}}}\over E_{{\bf k}}}
\right ){Z_{F}\over i\omega_{n}-E_{{\bf k}}}\nonumber \\
&+&{1\over 2}\left (1- {\bar{\xi_{{\bf k}}}\over E_{{\bf k}}}
\right ){Z_{F}\over i\omega_{n}+E_{{\bf k}}}, \\
\Im^{\dagger}(k)&=&-{1\over 2}{\bar{\Delta}_{hZ}({\bf k})\over
E_{{\bf k}}}\left ( {Z_{F}\over i\omega_{n}-E_{{\bf k}}}-{Z_{F}
\over i\omega_{n}+ E_{{\bf k}}}\right ),
\end{eqnarray}
\end{mathletters}
with $\bar{\xi_{{\bf k}}}=Z_{F}\xi_{{\bf k}}$, $\bar {\Delta}_{hZ}
({\bf k})=Z_{F}\bar{\Delta}_{h}({\bf k})$, and the dressed charge
carrier quasiparticle spectrum $E_{{\bf k}}=\sqrt{\bar
{\xi^{2}_{{\bf k}}}+\mid\bar{\Delta}_{hZ}({\bf k})\mid^{2}}$.

Although superconductivity with both d-wave and s-wave symmetries
appear within the $t$-$J$ model, the SC state has a dominated
d-wave symmetry in the optimal doping \cite{feng2}. Moreover, we
\cite{feng4} have discussed the effect of the additional second
neighbor hopping $t'$ on superconductivity, and found that the
d-wave SC pairing correlation is enhanced, while the s-wave SC
pairing correlation is heavily suppressed. In this paper, we are
interested in the charge asymmetry in superconductivity of hole-
and electron-doped cuprates. To make the discussions simpler, we
only consider the d-wave case, i.e., $\bar{\Delta}_{hZ}({\bf k})=
\bar{\Delta}_{hZ} \gamma^{(d)}_{{\bf k}}$, with
$\gamma^{(d)}_{{\bf k}}=({\rm cos} k_{x}-{\rm cos} k_{y})/2$. In
this case, the dressed charge carrier effective gap parameter and
single particle coherent weight in Eq. (4) satisfy the following
equations \cite{feng1,feng2},
\begin{mathletters}
\begin{eqnarray}
1&=&{1\over N^{3}}\sum_{{\bf k,q,p}}(Zt\gamma_{{\bf k+q}}-Zt'
\gamma_{{\bf k+q}}')^{2}\gamma^{(d)}_{{\bf k-p+q}}
\gamma^{(d)}_{{\bf k}}{Z^{2}_{F}\over  E_{{\bf k}}}{B_{{\bf q}}
B_{{\bf p}}\over\omega_{{\bf q}}\omega_{{\bf p}}}\nonumber\\
&\times&\left( {F^{(1)}_{1}({\bf k,q,p})\over (\omega_{{\bf p}}-
\omega_{{\bf q}})^{2}-E^{2}_{{\bf k}}}+{F^{(2)}_{1}({\bf k,q,p})
\over (\omega_{{\bf p}}+\omega_{{\bf q}})^{2}- E^{2}_{{\bf k}}}
\right ) ,\\
Z_{F}&=&1+{1\over N^{2}}\sum_{{\bf q,p}}(Zt\gamma_{{\bf p+k_{0}}}
-Zt'\gamma_{{\bf p+k_{0}}}')^{2}Z_{F}{B_{{\bf q}}B_{{\bf p}}\over
4\omega_{{\bf q}}\omega_{{\bf p}}}\nonumber \\
&\times&\left({F^{(1)}_{2}({\bf q,p}) \over (\omega_{{\bf p}}-
\omega_{{\bf q}}-E_{{\bf p-q+k_{0}}})^{2}}+{F^{(2)}_{2}({\bf q,p})
\over (\omega_{{\bf p}}-\omega_{{\bf q}}+E_{{\bf p-q+k_{0}}})^{2}}
\right . \nonumber \\
&+& \left . {F^{(3)}_{2}({\bf q,p} )\over (\omega_{{\bf p}}+
\omega_{{\bf q}}-E_{{\bf p-q+k_{0}}} )^{2}}+{F^{(4)}_{2}({\bf
q,p})\over (\omega_{{\bf p}}+\omega_{{\bf q}}+E_{{\bf p-q+k_{0}}}
)^{2}} \right ) ,
\end{eqnarray}
\end{mathletters}
respectively, where $F^{(1)}_{1}({\bf k,q,p})=(\omega_{{\bf p}}-
\omega_{{\bf q}})[n_{B}(\omega_{{\bf q}})-n_{B}(\omega_{{\bf p}})]
[1-2 n_{F}(E_{{\bf k}})]+E_{{\bf k}}[n_{B}(\omega_{{\bf p}})n_{B}(
-\omega_{{\bf q}})+n_{B}(\omega_{{\bf q}})n_{B}(-\omega_{{\bf p}})
]$, $F^{(2)}_{1}({\bf k,q,p})=-(\omega_{{\bf p }}+\omega_{{\bf
q}}) [n_{B}(\omega_{{\bf q}})-n_{B}(-\omega_{{\bf p}})][1-2 n_{F}
(E_{{\bf k}})]+E_{{\bf k}}[n_{B}(\omega_{{\bf p}}) n_{B}
(\omega_{{\bf q}})+n_{B}(-\omega_{{\bf p}})n_{B}(-\omega_{{\bf q}
})]$, $F^{(1)}_{2}({\bf q,p})=n_{F}(E_{{\bf p- q+k_{0}}})[n_{B}
(\omega_{{\bf q}})-n_{B}(\omega_{{\bf p}})]- n_{B}(\omega_{{\bf
p}})n_{B}(-\omega_{{\bf q}})$, $F^{(2)}_{2} ({\bf q,p})=
n_{F}(E_{{\bf p-q+k_{0}}}) [n_{B}(\omega_{{\bf p}})-n_{B}
(\omega_{{\bf q}})]-n_{B}(\omega_{{\bf q}})n_{B} (-\omega_{{\bf
p}})$, $F^{(3)}_{2}({\bf q,p})= n_{F}(E_{{\bf p-q+k_{0}}})
[n_{B}(\omega_{{\bf q}})-n_{B}(-\omega_{{\bf p}})]+n_{B}
(\omega_{{\bf p}})n_{B}(\omega_{{\bf q}})$, $F^{(4)}_{2}({\bf q,p}
)=n_{F}(E_{{\bf p-q+k_{0}}})[n_{B}(-\omega_{{\bf q}})-n_{B}
(\omega_{{\bf p}})]+n_{B}(-\omega_{{\bf p}})n_{B}(-\omega_{{\bf
q}})$, and $n_{B}(\omega)$ and $n_{F}(\omega)$ are the boson and
fermion distribution functions, respectively. These two equations
must be solved simultaneously with other self-consistent equations
\cite{feng1,feng2},
\begin{mathletters}
\begin{eqnarray}
\phi_{1}&=&{1\over 2N}\sum_{{\bf k}}\gamma_{{\bf k}}\left
(1-{\xi_{{\bf k}}\over E_{{\bf k}}}{\rm th}
[{1\over 2}\beta E_{{\bf k}}]\right ),\\
\phi_{2}&=&{1\over 2N}\sum_{{\bf k}}\gamma_{{\bf k}}'\left
(1-{\xi_{{\bf k}}\over E_{{\bf k}}}{\rm th}
[{1\over 2}\beta E_{{\bf k}}]\right ),\\
\delta &=& {1\over 2N}\sum_{{\bf k}}\left (1-{\xi_{{\bf k}} \over
E_{{\bf k}}}{\rm th}[{1\over 2}\beta E_{{\bf k}}]
\right ),\\
\chi_{1}&=&{1\over N}\sum_{{\bf k}}\gamma_{{\bf k}} {B_{{\bf
k}}\over 2\omega_{{\bf k}}}{\rm coth}
[{1\over 2}\beta\omega_{{\bf k}}],\\
\chi_{2}&=&{1\over N}\sum_{{\bf k}}\gamma_{{\bf k}}'{B_{{\bf
k}}\over 2\omega_{{\bf k}}}{\rm coth}
[{1\over 2}\beta\omega_{{\bf k}}],\\
C_{1}&=&{1\over N}\sum_{{\bf k}}\gamma^{2}_{{\bf k}} {B_{{\bf
k}}\over 2\omega_{{\bf k}}}{\rm coth}
[{1\over 2}\beta\omega_{{\bf k}}],\\
C_{2}&=&{1\over N}\sum_{{\bf k}}\gamma'^{2}_{{\bf k}} {B_{{\bf
k}}\over 2\omega_{{\bf k}}}{\rm coth}
[{1\over 2}\beta\omega_{{\bf k}}],\\
C_{3}&=&{1\over N}\sum_{{\bf k}}\gamma_{{\bf k}}\gamma_{{\bf k}}'
{B_{{\bf k}}\over 2\omega_{{\bf k}}}{\rm coth}
[{1\over 2}\beta\omega_{{\bf k}}],\\
{1\over 2} &=&{1\over N}\sum_{{\bf k}}{B_{{\bf k}} \over
2\omega_{{\bf k}}}{\rm coth}
[{1\over 2}\beta\omega_{{\bf k}}],\\
\chi^{z}_{1}&=&{1\over N}\sum_{{\bf k}}\gamma_{{\bf k}}
{B_{z}({\bf k})\over 2\omega_{z}({\bf k})}{\rm coth}
[{1\over 2}\beta\omega_{z}({\bf k})],\\
\chi^{z}_{2}&=&{1\over N}\sum_{{\bf k}}\gamma_{{\bf k}}'
{B_{z}({\bf k})\over 2\omega_{z}({\bf k})}{\rm coth}
[{1\over 2}\beta\omega_{z}({\bf k})],\\
C^{z}_{1}&=&{1\over N}\sum_{{\bf k}}\gamma^{2}_{{\bf k}}
{B_{z}({\bf k})\over 2\omega_{z}({\bf k})}{\rm coth} [{1\over
2}\beta\omega_{z}({\bf k})],\\
C^{z}_{3}&=&{1\over N}\sum_{{\bf k}}\gamma_{{\bf k}}\gamma_{{\bf
k}}'{B_{z}({\bf k})\over 2\omega_{z}({\bf k})}{\rm coth} [{1\over
2}\beta\omega_{z}({\bf k})],
\end{eqnarray}
\end{mathletters}
then all the above order parameters, decoupling parameter
$\alpha$, and chemical potential $\mu$ are determined by the
self-consistent calculation \cite{feng1,feng2}.

It has been shown \cite{feng1} that the dressed charge carrier
pairing state originating from the kinetic energy terms by
exchanging dressed spin excitations can lead to form the electron
Cooper pairing state, where the SC gap function is obtained from
the electron off-diagonal Green's function $\Gamma^{\dagger}(i-j,
t-t')=\langle\langle C^{\dagger}_{i\uparrow}(t);
C^{\dagger}_{j\downarrow}(t')\rangle\rangle$, which is a
convolution of the dressed spin Green's function and dressed
charge carrier off-diagonal Green's function, and reflects the
charge-spin recombination \cite{anderson1}. In the present case,
this electron off-diagonal Green's function can be obtained in
terms of the MF dressed spin Green's function (5) and dressed
charge carrier off-diagonal Green's function (6b) as,
\begin{eqnarray}
\Gamma^{\dagger}(k)&=&{1\over N}\sum_{{\bf p}}{Z_{F}
\bar{\Delta}_{hZ}({\bf p-k})\over E_{{\bf p-k}}}{B_{{\bf p}}\over
2\omega_{{\bf p}}} \nonumber \\
&\times& \left ({(\omega_{{\bf p}}+E_{{\bf p-k}})
[n_{B}(\omega_{{\bf p}})+n_{F}(-E_{{\bf p-k}})]\over
(i\omega_{n})^{2}-(\omega_{{\bf p}}+E_{{\bf p-k}})^{2}}
\right . \nonumber \\
&-&\left .{(\omega_{{\bf p}}-E_{{\bf p-k}})[n_{B}(\omega_{{\bf p}}
)+n_{F}(E_{{\bf p-k}})]\over (i\omega_{n})^{2}-(\omega_{{\bf p}}-
E_{{\bf p-k}})^{2}}\right ).
\end{eqnarray}
With the help of this electron off-diagonal Green's function, the
SC gap function is obtained as $\Delta({\bf k})=-(1/\beta)
\sum_{i\omega_{n}}\Gamma^{\dagger}({\bf k},i\omega_{n})$, and can
be evaluated explicitly,
\begin{eqnarray}
\Delta({\bf k})&=&-{1\over N}\sum_{{\bf p}}{Z_{F}\bar{\Delta}_{hZ}
({\bf p-k})\over 2E_{{\bf p-k}}} \nonumber \\
&\times&{\rm tanh} [{1\over 2}\beta E_{{\bf p-k}}]{B_{{\bf p}}
\over 2\omega_{{\bf p} }}{\rm coth}[{1\over 2}\beta\omega_{{\bf
p}}],
\end{eqnarray}
which shows that the SC transition temperature $T_{c}$ occurring
in the case of the SC gap parameter $\Delta=0$ is identical to the
dressed charge carrier pair transition temperature occurring in
the case of the effective dressed charge carrier pairing gap
parameter $\bar{\Delta}_{hZ}=0$. Since the absolute values of $t$
and $t'$ are almost same for both hole- and electron-doped
cuprates \cite{hybertson}, and therefore in this paper, the
commonly used parameters are chosen as $t/J=2.5$ and $t'/t=0.3$
for the hole doping, and $t/J=-2.5$ and $t'/t=0.3$ for the
electron doping. In Fig. 1, we plot the SC transition temperature
$T_{c}$ as a function of the doping concentration $\delta$ for (a)
the electron doping and (b) the hole doping in comparison with the
corresponding experimental results of
Pr$_{2-x}$Ce$_{x}$CuO$_{4-y}$ \cite{peng} and
La$_{2-x}$Sr$_{x}$CuO$_{4}$ \cite{tallon} (inset). Our results
indicate that for the hole-doped case, superconductivity appears
over a wide range of doping, where the maximal SC transition
temperature T$_{c}$ occurs around the optimal doping concentration
$\delta_{{\rm opt}}\approx 0.15$, and then decreases in both
underdoped and overdoped regimes. In analogy to the phase diagram
of the hole-doped case, superconductivity appears over a narrow
range of doping in the electron-doped side, where the SC
transition temperature T$_{c}$ increases sharply with increasing
doping in the underdoped regime, and reaches a maximum in the
optimal doping $\delta_{{\rm opt}}\approx 0.14$, then decreases
sharply with increasing doping in the overdoped regime. However,
the maximum achievable SC transition temperature in the optimal
doping in electron-doped cuprates is much lower than that of the
hole-doped case due to the electron-hole asymmetry. Using an
reasonably estimative value of $J\sim 800$K to 1200K in doped
cuprates, the SC transition temperature in the optimal doping is
T$_{c}\approx 0.22J \approx 176{\rm K}\sim 264{\rm K}$ for the
hole-doped case, and T$_{c}\approx 0.136J \approx 108{\rm K}\sim
163{\rm K}$ for the electron-doped case, in qualitative agreement
with the corresponding experimental data
\cite{tallon,tokura,peng}.

\begin{figure}[prb]
\epsfxsize=3.5in\centerline{\epsffile{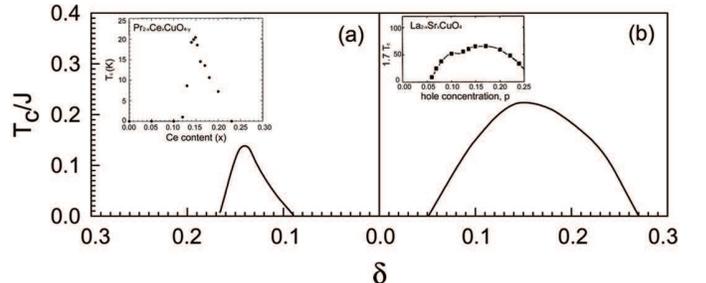}}\caption{The
superconducting transition temperature as a function of the doping
concentration with (a) $t/J=-2.5$ and $t'/t=0.3$ for the electron
doping and (b) $t/J=2.5$ and $t'/t=0.3$ for the hole doping.
Inset: the corresponding experimental results of
Pr$_{2-x}$Ce$_{x}$CuO$_{4-y}$ taken from Ref. [8] and
La$_{2-x}$Sr$_{x}$CuO$_{4}$ from Ref. [3].}
\end{figure}

The essential physics of the doping dependent SC transition
temperature in the electron-doped case is almost the same as in
the hole-doped side, and detailed explanations have been given in
Ref. [18]. In the framework of the kinetic energy driven
superconductivity \cite{feng1}, the self-energy function
$\Sigma^{(h)}_{2}({\bf k})$ describes the effective dressed charge
carrier pairing gap function, and measures the strength of the
binding of charge carrier pairs, while the antisymmetric part of
the self-energy function $\Sigma^{(h)}_{1o} ({\bf k})$ (then
$Z_{F}$) describes the single particle (quasiparticle) coherence,
and therefore $Z_{F}$ is closely related to the quasiparticle
density. Since the SC state is established through an emerging
quasiparticle \cite{ding}, then the SC state is controlled by both
gap function and quasiparticle coherence, which is reflected
explicitly in the self-consistent equations (7a) and (7b). It has
been shown that the doping dependent behavior of the single
particle coherent weight resembles that of the superfluid density
in doped cuprates \cite{feng2}, i.e., $Z_{F}$ grows linearly with
the doping concentration in the underdoped and optimally doped
regimes, and then decreases with increasing doping in the
overdoped regime, which leads to that the SC transition
temperature reaches a maximum in the optimal doping, and then
decreases in both underdoped and overdoped regimes. On the other
hand, it has been shown \cite{hybertson} that AFLRO can be
stabilized by the $t'$ term for the electron-doped case, which may
lead to the charge carrier's localization over a broader range of
doping, this is also why superconductivity appears over a narrow
range of doping in electron-doped cuprates.

In the CSS fermion-spin theory \cite{feng3}, the physical electron
is decoupled as the dressed charge carrier $h_{i\sigma}=
e^{-i\Phi_{i\sigma}}h_{i}$ and spin $S_{i}$. Since the phase
factor $\Phi_{i\sigma}$ in the dressed charge carrier is separated
from the bare spinon operator, and then it describes a spin cloud
\cite{feng3}. Therefore the dressed charge carrier $h_{i\sigma}$
is a spinless fermion $h_{i}$ (bare charge carrier) incorporated
the spin cloud $e^{-i\Phi_{i\sigma}}$ (magnetic flux), thus is a
magnetic dressing. In other words, the dressed charge carrier
carries some spin messages, i.e., it shares its nontrivial spin
environment. It has been shown \cite{feng3} that these dressed
charge carrier and spin are gauge invariant under a local U(1)
gauge transformation, and in this sense, they are real and can be
interpreted as the physical excitations. In doped cuprates, the
normal-state above the SC transition temperature exhibits a number
of anomalous properties which is due to CSS \cite{anderson}, while
the SC state is characterized by the charge-spin recombination
\cite{anderson1}. Based on the CSS fermion-spin theory, we
\cite{feng6} have discussed the charge dynamics of the underdoped
cuprates in the normal-state, and show that under temperature
$T^{*}$, the magnetic fluctuation is strong enough to lead to a
pseudogap. This pseudogap would reduce the charge carrier
scattering and thus is responsible for the temperature linear to
the nonlinear range in the in-plane resistivity and the crossovers
to the insulating-like range in the c-axis resistivity.
Furthermore, the temperature $T^{*}$ is doping dependent, and
grows monotonously as the doping concentration decreases, and
disappear in higher doping \cite{feng6}. It has been shown
\cite{feng6} that this pseudogap (then the temperature $T^{*}$) is
obtained from the charge carrier Green's function in the
normal-state by considering the second-order correction due to the
spin pair bubble. In the kinetic energy driven SC mechanism
\cite{feng1,feng2}, the charge carrier pairing state (then the
electron SC-state and SC transition temperature $T_{c}$) occurrs
directly through the kinetic energy by exchanging spin
excitations, and is controlled by both charge carrier gap function
and single particle coherent weight. This single particle coherent
weight is obtained from the charge carrier quasiparticle diagonal
Green's function in the SC-state by considering the second-order
correction due to the spin pair bubble, and at the same time is
effected by the dressed charge carrier pair gap function (then the
charge carrier quasiparticle off-diagonal Green's function), which
is shown explicitly in the self-consistent equations (7a) and
(7b). Moreover, this dressed charge carrier pairing amplitude
(then the SC order parameter) is proportional to the number of
charge carrier quasiparticles (then the superfluid density), and
not to the number of electrons as mentioned above, which leads to
\cite{feng2} that the single particle coherent weight
$Z_{F}(T_{c})$ resembles that of the superfluid density. In other
words, $T^{*}$ is closely related to the spin fluctuation, while
$T_{c}$ is self-consistently governed by the single particle
coherence and dressed charge carrier pair gap function, this is
why there are some differences between $T^{*}$ and $T_{c}$.

In summary, within the framework of the kinetic energy driven the
SC mechanism \cite{feng1}, we have discussed the charge asymmetry
in superconductivity of hole- and electron-doped cuprates based on
the $t$-$t'$-$J$ model. Our results show that for the hole-doped
case, superconductivity appears over a wide range of doping, where
the maximal SC transition temperature occurs around the optimal
doping concentration, and then decreases in both underdoped and
overdoped regimes. In analogy to the phase diagram of hole-doped
case, superconductivity appears over a narrow range of doping in
the electron-doped side, where the SC transition temperature
increases sharply with increasing doping in the underdoped regime,
and reaches a maximum in the optimal doping, then decreases
sharply with increasing doping in the overdoped regime. However,
the maximum achievable SC transition temperature in the optimal
doping in the electron-doped case is much lower than that of the
hole-doped side due to the electron-hole asymmetry. Our these
results are in qualitative agreement with the experimental
observations.

\acknowledgments

The author would like to thank Dr. Huaiming Guo for the helpful
discussions. This work was supported by the National Natural
Science Foundation of China under Grant Nos. 10125415 and
90403005, and the Grant from Beijing Normal University.

\end{document}